\documentclass[superscriptaddress,twocolumn,aps,prb,ltxgrid,showpacs]{revtex4}
\usepackage{epsfig}
\usepackage{graphicx}
\usepackage{amsbsy,amssymb,amsmath,bm,ulem}

\usepackage[usenames]{color}

\begin{document}

\title{Flux Distribution in Superconducting Films with Holes}
\author{J. I. Vestg\aa rden, D. V. Shantsev, Y. M. Galperin and T. H. Johansen}
\affiliation{Department of Physics and Center for Advanced
Materials and Nanotechnology,    University of Oslo, P. O. Box
1048 Blindern, 0316 Oslo, Norway}

\begin{abstract}
Flux penetration into type-II superconducing films is 
simulated for transverse applied magnetic field
and flux creep dynamics.
The films contain macroscopic, non-conducting holes 
and we suggest a new method to introduce the holes in
the simulation formalism. The method implies 
reconstruction of the magnetic field change inside the hole.
We find that in the region between the hole and the edge
the current density is compressed so that the flux density is slightly reduced,
but the traffic of flux is significantly increased.
The results are in good agreement
with magneto-optical studies of flux distributions in YBa$_2$Cu$_3$O$_x$ films.
\end{abstract}

\maketitle

\section{Introduction}

The behavior of vortex matter in superconductors can 
to a large degree be controlled by  
introducing artificial defects.
It has been known for a long time that 
randomly distributed defects, created e.g. by neutron irradiation, 
allow a dramatic enhancement of
the critical current density, $j_c$.
One may reach more specific goals by 
tuning the arrangement of artificial defects.
In particular, experiments on superconducting thin films 
have revealed a large number of interesting 
effects,  including
matching effects,\cite{moshchalkov98} noise reduction in SQUIDs,\cite{wordenweber02} 
rectified vortex motion,\cite{desilva06,vondel05} 
anisotropy of $j_c$,\cite{pannetier03} 
and vortex guidance.\cite{wordenweber04}  

In parallel with the experimental progress, 
the theoretical understanding of 
how artificially created patterns interact with vortex matter 
is also developing. 
Interaction between a single vortex and a cylindrical cavity 
in a bulk superconductor was considered 
within the London approximation in Ref.~\onlinecite{nordborg00}.
This work extends the classical paper, Ref.~\onlinecite{mkrtchyan72}, 
predicting the maximal number of flux quanta
that can be trapped by a single hole.
Current distribution around a 1D array of 
holes was calculated within the Ginsburg-Landau theory 
in Ref.~\onlinecite{wordenweber04}.
However, these theoretical works   
consider a {\em bulk} superconductor, while
most experiments are on patterned 
{\em thin films}.\cite{moshchalkov98,wordenweber02,desilva06,vondel05,pannetier03,wordenweber04}
Moreover, a realistic model should 
take into account the strong pinning
of vortices in the superconducting areas around the artificial defects.
When the defect size is much larger than the London penetration depth,
one can consider the average vortex density $B$ rather than individual vortices.
Such an approach was used in Ref.~\onlinecite{gheorghe06}
to simulate flux penetration into a thin film with a 2D array of holes.
It allowed to explain an asymmetrical flux penetration
due to asymmetry in the hole shape.  
At the same time, the case of an individual hole in a thin film has not yet
been carefully analysed. A main purpose of the present work 
is to acquire details of flux and current distributions 
in a superconducting strip with one individual hole.

An approximate picture of the current distribution 
around a non-conducting hole can be obtained within 
Bean's critical state model.\cite{bean64}
In the Bean model current stream lines are added from 
the edge with equal spacing representing the critical
current density. The presence of a hole forces 
the current to flow around it
and hence pushes the flux front deeper into the sample.
Both holes and sample corners give rise to so-called
$d$-lines where the current changes direction
discontinuously.\cite{CCS} 
They are seen as dark lines\cite{note:mo-dark-bright}
in images showing magnetic flux distributions.\cite{schuster94,jooss02}
For example, 90$^\circ$~corners give 45$^\circ$~straight $d$-lines\cite{brandt95} 
while semicircular indentations of the edge 
give parabolic $d$-lines.\cite{mints96}
The magneto-optical image of Fig.~\ref{fig:mo-hole}
shows $d$-lines spreading out from a circular hole
towards the flux-free region.
The same hole also introduces another pattern: 
a darkened region starting from the 
hole and extending {\it towards the edge}.
This pattern is similar to the one observed by 
Eisenmenger et al., Ref.~\onlinecite{eisenmenger01}.
The pattern does not fit with the common
interpretation of the Bean model, which leaves the 
the currents between the hole and the edge unperturbed.
Ref.~\onlinecite{eisenmenger01} 
discusses how to reinterpret
the Bean model and explain the
observed pattern as a second parabolic $d$-line.
In this work, we will go further and do   
full dynamical simulations
of flux penetration taking into account
the  non-local electrodynamics of films as well as
flux creep. Our results provide details of flux 
and current distributions in the vicinity of a hole and 
suggest a new interpretation for the observed anomaly.

\section{Model}
\subsection{Single-connected superconductors}

Consider a type-II superconducting thin film placed in an 
increasing transverse magnetic field. 
The superconductor responds by 
generating screening currents to  shield its interior. 
The current density is highest at the edges where the Lorentz force 
eventually overcomes the pinning force, leading to penetration of flux.
According to the Bean model, the vortices move only when 
the local current density exceeds the critical value, $j_c$.
A more realistic model for flux penetration also allows  
for flux creep at $j<j_c$. 
Macroscopically, flux creep 
is introduced through a highly non-linear current
voltage relation\cite{zeldov90,brandt95}
\begin{equation}
  \mathbf E =  \rho_0\left(\frac{j}{j_c}\right)^{n-1}\mathbf j
  ,
  \label{material-law}
\end{equation}
where $\mathbf E$ is electric field, $\rho_0$ a resistivity constant,
$\mathbf j$ is current density, 
and $n$ is the creep exponent. For thin films of 
YBa$_2$Cu$_3$O$_x$, $n$ is typically in the range from 10 to 70
depending on temperature and pinning strength.\cite{sun91}

\begin{figure}[t]
  \centering
  \epsfig{file=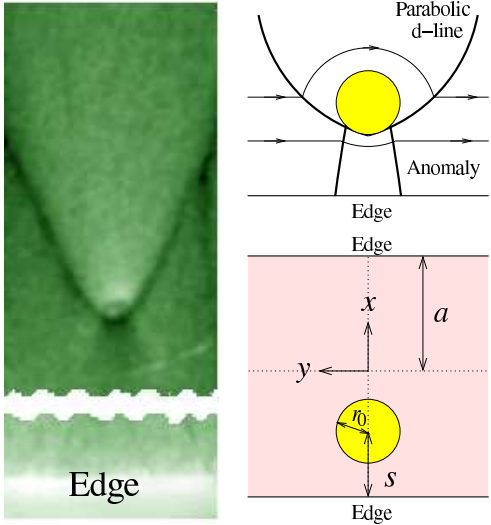, height=7cm}
  \caption{
    Left: Magneto-optical image of $B_z$ near a hole. 
    Note the parabolic $d$-lines going upwards and a dark area 
    going downwards from the hole. 
    Right: a sketch of the strip with a 
    circular hole indicating how peculiarities in the flux distribution 
    are related to bending of the current stream-lines.
    Notations: film half-width is
    $a$, distance from the edge to the hole center is $s$, and the  
    hole radius is $r_0$.}
  \label{fig:mo-hole}
  \label{fig:sample}
\end{figure}

Flux dynamics of single-connected type-II superconductors 
in transverse geometry 
has been described thoroughly by E.~H.~Brandt.
This work uses the same formalism and hence we 
only give a short summary of the simulation basics, 
mainly following Refs.~\onlinecite{brandt95-prl,brandt95}
and \onlinecite{brandt05}. The next section
will be devoted to additional changes for multiply-connected samples.

For films, it is a great simplification to work with the sheet current
$\mathbf J(\mathbf r)=\int_{-d/2}^{d/2} dz~\mathbf j(\mathbf r,z)$, 
$\mathbf r=(x,y)$,
in stead of the current density $\mathbf j$. 
This is justified as long as thickness, $d$, is small compared to 
the in-plane dimensions but much larger
than the London penetration depth, $\lambda$. Finite $\lambda$
can be handled with a small modification of the algorithm.\cite{brandt05}
Since the current is conserved, $\nabla\cdot \mathbf J=0$, 
it can be expressed as $\mathbf J=\nabla\times \hat zg$,
where $g=g(\mathbf r)$ is the local magnetization.\cite{brandt95-prl} 

For single-connected thin films the Biot-Savart 
law can be formulated as  
\begin{equation}
  B_z(\mathbf r,z)/\mu_0
  = H_a+
  \int_{A} d^2 r'
  ~Q(\mathbf r, \mathbf r',z)~g(\mathbf r') 
  ,
  \label{hfromg}
\end{equation}
where $H_a$ is the applied magnetic field, and 
$A$ is the sample area. The kernel $Q$ represents 
the field generated by a dipole of unit strength,\cite{brandt95}
\begin{equation}
 Q(\mathbf r,\mathbf r', z) = \frac{1}{4\pi}
  \frac{2z^2-(\mathbf r - \mathbf r')^2}
       {\left[z^2+(\mathbf r-\mathbf r')^2\right]^{5/2}}\,.
\end{equation}
We discretize the kernel on an equidistant grid with grid points $r_i$ 
and weights $w$ and obtain\cite{brandt05}
\begin{equation}
  Q_{ij}
  =
    \delta_{ij}\left(C_i/w+\sum_l q_{il}\right)
    -
    q_{ij}    
  ,
  \label{kernel-discrete}
\end{equation}
where $q_{ij}= 1/4\pi|\mathbf r_i - \mathbf r_j|^3$ for $i\neq j$ and $q_{ii}=0$. 
The function $C$ depends on the sample geometry. It is given as
\begin{equation}
  \label{C}
  C(\mathbf r) 
  =  
  \int_\text{outside}
  \frac{dr'^2}
       {4\pi |\mathbf r - \mathbf r'|^3}
       .
\end{equation}

The time evolution of $g$ comes from the inverse of 
Eq.~\eqref{hfromg},
\begin{equation}
  \dot g(\mathbf r) 
  =
  \int_A d^2 r'~Q^{-1}(\mathbf r,\mathbf r')~[\dot B_z(\mathbf r')-\mu_0\dot H_a]
  ,
  \label{dynamics}
\end{equation}
where $Q^{-1}$ for discrete problem is the matrix inverse of Eq.~\eqref{kernel-discrete}.
$\dot B_z$ is given from Faraday's law as
\begin{equation}
  \dot B_z(\mathbf r)=-\left(\nabla\times\mathbf E\right)_z=\nabla\cdot(\frac{\rho}{d\mu_0}\nabla g)
  ,
  \label{B_z}
\end{equation}
with $\rho= \rho_0|\nabla g/J_c|^{n-1}$ obtained from Eq.~\ref{material-law}. 
The right-hand side of Eq.~\eqref{dynamics} is expressed only via $g$ and $H_a$
so that time evolution of $g$ can be found by integrating the equation numerically.

\begin{figure}[t]
  \centering
  \epsfig{file=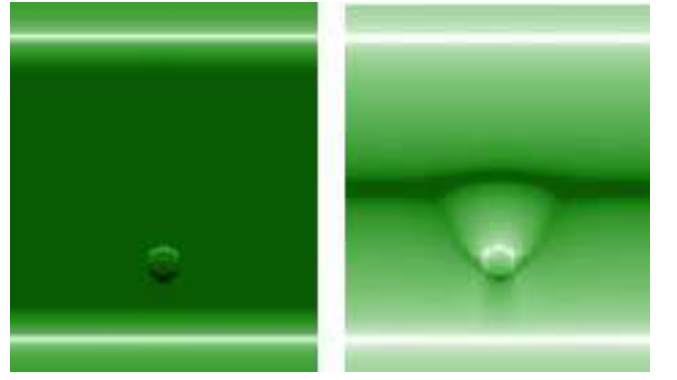, width=8.5cm}
  \caption{Simulated magnetic field distribution 
    in a long strip, plotted in the style 
    of magneto-optical images, where the intensity represents $B_z$. 
    At small applied field (left) the hole produces a field dipole and 
    at large field (right) one can see the parabolic $d$-lines and
    a dark region between the hole and the edge, 
    cf. the experimental image, Fig.~\ref{fig:mo-hole}; 
    $r_0/a=0.1$, $s/a=0.5$, $n=19$,  $H_a/J_c=0.2$ (left) and 1 (right), and $\mu_0\dot H_a=\rho_0J_c/ad$.
  }
   \label{fig:B-full}
   \label{fig:B-half}
\end{figure}

\subsection{Superconductors with holes}
\label{sec:boundary-holes}

For macroscopic, arbitrarily shaped, single-connected, 
type-II superconducting films flux 
dynamics is fully described by Eq.~\eqref{dynamics}.
This basic equation can also be used for multiply connected 
samples, but in this case one needs to specify the 
dynamically changing value of
$g$ at the hole boundary. 
In Refs.~\onlinecite{gheorghe06} and \onlinecite{loerincz04}
this value was set to the lowest value of $g$ 
along the hole perimeter. This method
turned out to be quite feasible 
but unfortunately it cannot reproduce the discussed pattern of 
Fig.~\ref{fig:mo-hole}. Moreover, it also 
introduces unphysical net flux into the hole before 
the flux front has reached it.

A completely different approach is
to consider the holes as part of the sample, but  
ascribe to them a large Ohmic resistance 
or a strongly reduced $J_c$.\cite{crisan2005}
Then, Eq.~\eqref{dynamics} applies to the whole sample including the holes,
while the material law, Eqs.~\eqref{material-law} and \eqref{B_z},
is spatially non-uniform.
This approach is physically justified
but numerically challenging due to 
huge electric field gradients. In addition,
there still remain small but non-zero currents flowing
within the holes.

\begin{figure}[t]
  \centering
  \epsfig{file=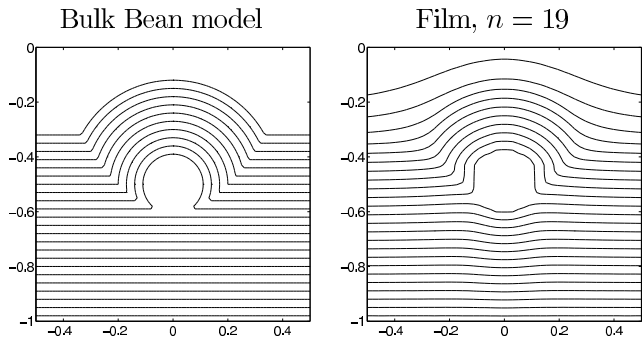, width=8.5cm}
  \caption{The current stream lines for the bulk Bean model (left)
    and for film with finite $n$ (right). 
    }
  \label{fig:j-stream}
\end{figure}

In this work we propose a new approach
that does not require any additional assumptions, though requires
a larger computational time. 
In this approach the integration in Eq.~\eqref{dynamics} 
is extended over the whole sample area including the holes.
Then the dynamics of $g$ is described by the equation
\begin{equation}
  \dot g(\mathbf r) 
  =
  \int_A d^2 r'~Q^{-1}(\mathbf r,\mathbf r')~
      [\dot B^{(s)}_z(\mathbf r')+\dot B_z^{(h)}(\mathbf r')-\mu_0\dot H_a]
  ,
  \label{dynamics-hole}
\end{equation}
where $A$ is the sample area including the hole.
Here we presented $\dot B_z$ as a sum $\dot B_z^{(h)}+\dot B_z^{(s)}$
where $\dot B_z^{(h)}$ is nonzero only in the hole 
and  $\dot B_z^{(s)}$ is nonzero within the superconducting areas. 
$\dot B_z^{(s)}$ is calculated in the straightforward way using Eq.~\eqref{B_z}.
The other term, $\dot B_z^{(h)}$, is defined by two conditions. 
The first condition is that current does not flow beyond the superconducting areas, i.e.,  
$\dot g$ is constant within the hole.
This constant is determined by the  
second condition, that the total change of magnetic flux inside the hole
is related to the electric field 
at its boundary through Faraday's law,
\begin{equation}
  \int_\text{hole} d^2r~\dot B_z = -\int_\text{hole edge}d\mathbf l\cdot \mathbf E
  .
  \label{constraint-hole}
\end{equation}

\begin{figure}[t]
  \centering
  \epsfig{file=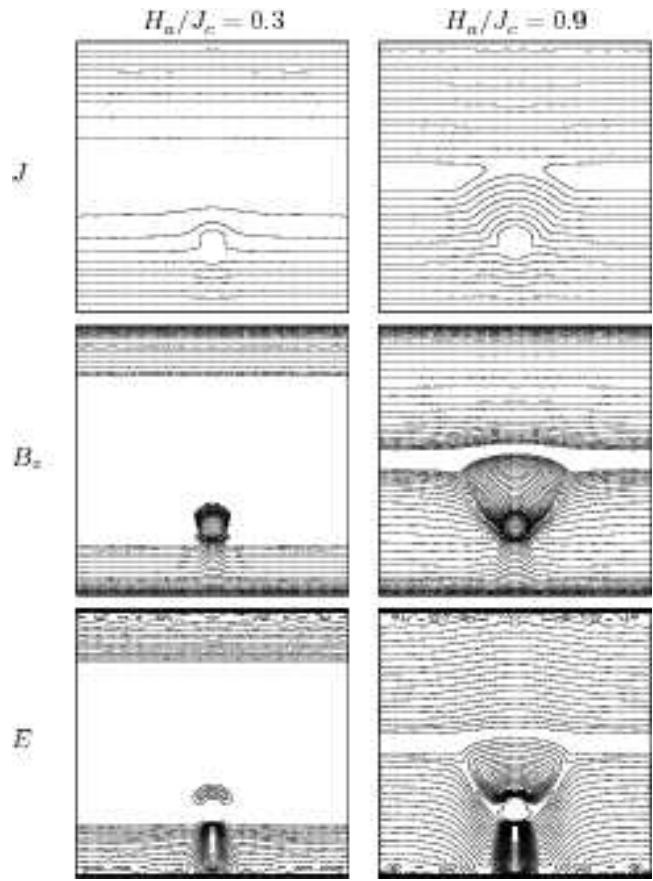, width=8.5cm}
  \caption{Simulation results for a strip with a hole:  the 
    current stream lines (top), $B_z$ contour lines (middle), 
    and $E$ contour lines (bottom).
    Note that the electric field is greatly enhanced 
    in the channel between the hole and the edge;\cite{note:E-in-hole}
    $H_a/J_c=0.3$~and 0.9.
    The remaining parameters are the same as for Fig.~\ref{fig:B-half}. 
  }
  \label{fig:strip-jHE}
\end{figure}

In order to find a $\dot B_z^{(h)}$ that satisfies the two conditions
we use an iteration scheme.
An initial guess, $\dot B_z^{(h,0)}$, 
is substituted into Eq.~\eqref{dynamics-hole} to find $\dot g^{(h,0)}$ 
inside the hole. The next approximation is found as
\begin{equation}
  \dot B_z^{(h,1)}(\mathbf r) = \dot B_z^{(h,0)}(\mathbf r)-\int_\text{hole}d^2r' 
  Q(\mathbf r,\mathbf r') \dot g^{(h,0)}(\mathbf r') + K,
  \label{iteration-step}
\end{equation}
where the constant $K$ is chosen so that Eq.~\eqref{constraint-hole} is satisfied.
$\dot B_z^{(h,1)}$ is then inserted into Eq.~\eqref{dynamics-hole} 
to find  $\dot g^{(h,1)}$.
This $\dot g^{(h,1)}$ is in general non-uniform, but 
when the procedure is repeated $\dot g^{(h,n)}$ 
becomes more uniform with every new iteration.
A smart choice of the initial guess of $\dot B_z^{(h,0)}$
is the final value at the previous time step,
$\dot B_z^{(h,0)}(\mathbf r,t)=\dot B_z^{(h,n)}(\mathbf r,t-\Delta t)$.
With this choice only a couple of  
iterations are sufficient. 

Note that the scheme presented here is in no way bound to the 
discrete formulation of the kernel, Eq.~\eqref{kernel-discrete}.
It can be used for any formulation as long as both 
the forward and inverse relations between $\dot g$ and $\dot B_z$ are known.
Further mathematical details are in appendix~\ref{sec:numerics}.

\begin{figure}[t]
  \centering
  \epsfig{file=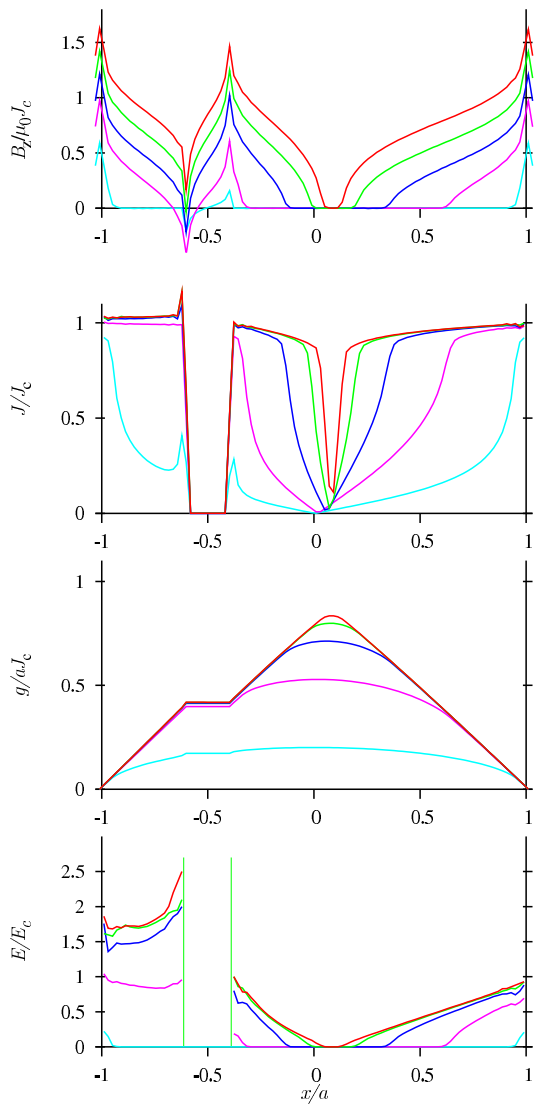, width=7cm}
  \caption{Profiles of $B_z$,  $J$, $g$ and $E$ 
    through $y=0$ for a strip with a hole.
    The curves correspond to applied fields
    $H_a/J_c = $ 0.1, 0.3, 0.5, 0.7, and 0.9, and $E_c=\rho_0J_c/d$.
    The remaining parameters are the same as for Fig.~\ref{fig:B-half}. 
  }
  \label{fig:strip-slices}
  \label{fig:H-slices}
  \label{fig:j-slices}
\end{figure}

\begin{figure}[t]
  \centering
  \epsfig{file=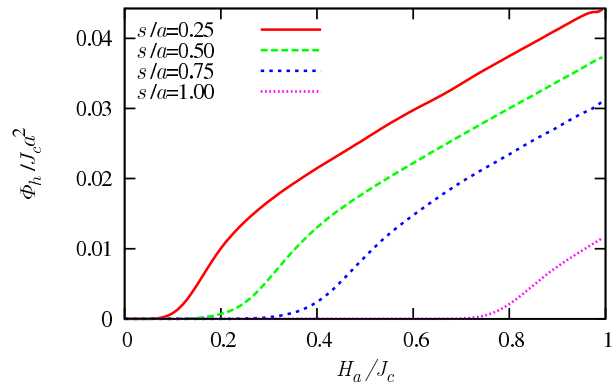, width=8cm}
  \caption{Total flux inside the hole, $\Phi_h=\int_\text{hole}d^2r~B_z$, 
    as a function of $H_a$, for 
    various distances $s$ from the edge. 
    For low fields $\Phi_h$ is zero since the flux front has not reached the hole yet.
    For high fields $\Phi_h$ grows linearly with $H_a$ since the strip is saturated with $J\approx J_c$.
    Hole radii are $r_0/a=0.1$.
    The remaining parameters are the same as for Fig.~\ref{fig:B-half}.
  }
  \label{fig:flux-hole}
\end{figure}

\section{Strip with a circular hole}
\label{sec:one-hole}

In this section, Eq.~\eqref{dynamics-hole} is solved for 
an infinite superconducting strip in linearly increasing magnetic field.
The strip is modeled using periodic boundary conditions in the $y$-direction,
and examples of magnetic field distributions are given in Fig.~\ref{fig:B-half}.
In the upper part one observes  regular flux penetration 
with maximum of $B_z$ at the edges. 
Flux penetration in the lower half
is strongly affected by the presence a small, circular, non-conducting hole.
Note that the flux distribution is perturbed 
in a region that significantly exceeds the hole dimensions.

The left image of Fig.~\ref{fig:B-half} corresponds to a small field for which the flux front
has not reached the hole yet. 
In this case the hole shows up as a field dipole, in agreement
with magneto-optical observations; cf. Refs.~\onlinecite{eisenmenger01}
and \onlinecite{yurchenko06}.
Namely, there is positive field at the farther side of the hole
and a negative field at the side closer to the film edge.
The negative fields shrinks when the flux front reaches the hole,
but the asymmetry of the flux distribution inside the hole
remains, as seen in the right image.
As expected, the front becomes distorted so that the
penetration is significantly deeper 
in the vicinity of the hole.
For the full penetration image of Fig.~\ref{fig:B-full} 
one also clearly see the 
$d$-lines as dark line originating at the hole
and directed towards the middle of the strip.
Such $d$-lines were first described in Ref.~\onlinecite{CCS}
within the Bean model framework
and they are called $d$-lines because 
current changes direction discontinuously there. 
The discontinuity is most clearly seen in 
current stream line plot of Fig.~\ref{fig:j-stream}~(left).
For the Bean model, $d$-lines from circular holes are parabolic 
and by convention $d$-lines from small holes inside superconductors 
are often called parabolas.
In the presence of flux creep 
the change of current direction is smeared as follows from Fig.~\ref{fig:j-stream}~(right).
However, the $d$-lines are still clearly visible, at least for $n\gg 1$.

Comparing the two panels of Fig.~\ref{fig:j-stream} we notice
a qualitative difference between the current flow in the bulk Bean model
and for films under the creep.
In the Bean model the current density is everywhere constant
and all the current that is blocked by the hole turns
towards the strip center. The region between
the hole and the edge is hence unaffected by the presence of the hole.
For film creep dynamics this is no longer true and
a certain fraction of the current will force its way here.
As a result, the current density is enhanced which is seen as 
denser stream lines in Fig.~\ref{fig:j-stream}~(right).
Since the stream lines bend they create the
feature visible in the flux distribution of Fig.~\ref{fig:B-half}:
a slightly darkened region starting at the hole and widening towards the edge. 
This feature can also be observed experimentally; cf. Fig.~\ref{fig:mo-hole}.
It was analysed in detail in Ref.~\onlinecite{eisenmenger01} and
interpreted in terms of the Bean model as additional parabolic d-lines.     
Our experiment and simulations suggest a different interpretation.
We believe that one should speak about  
an {\em area} of reduced flux density rather than new d-lines.
Moreover, the appearance of this area is due to {\em locally enhanced} current density,
hence it cannot be explained within the Bean model, postulating $J=J_c$.
An enhanced current density also implies a strongly enhanced electric field.
This is clearly seen in Fig.~\ref{fig:strip-jHE} showing 
the contour lines of $E$.
A locally enhanced $E$ means that there is an exceptionally intensive traffic of magnetic flux  
through the channel between the edge and the hole.
The channel width is approximately given by the hole diameter, but 
increases slightly towards the edge. The width depends in general
on the distance to the edge and the creep exponent $n$.
Both larger distance and smaller $n$ tend to make the channel wider.

After arrival to the hole, the flux is further directed
in the fan-shaped region between the d-lines. Electric field within
this region is also relatively high, again implying an intensive flux traffic.
This situation is similar to the case of a semicircular indentation 
at the sample edge considered in Refs.~\onlinecite{mints96}, \onlinecite{schuster96}, 
and \onlinecite{vestgarden07-2}.
The hole thus strongly rearranges trajectories of flux flow.

The above discussion is further confirmed by profiles 
of $B_z$, $J$, $g$, and $E$ through the line $y=0$
shown in Fig.~\ref{fig:strip-slices}.
The $J$ profiles show features commonly observed in strips,\cite{johansen96} i.e.,
plateaus with values $\sim J_c$ in the penetrated regions and shielding 
currents with $J<J_c$ in the Meissner regions. 
The profiles show clearly the  enhanced $J$ and $E$ between the edge and the hole.
It is also interesting to see the negative $B_z$ for 
low values and how the negative values gradually vanish
when the main flux front gets in contact.

\begin{figure}[t]
  \centering
  \epsfig{file=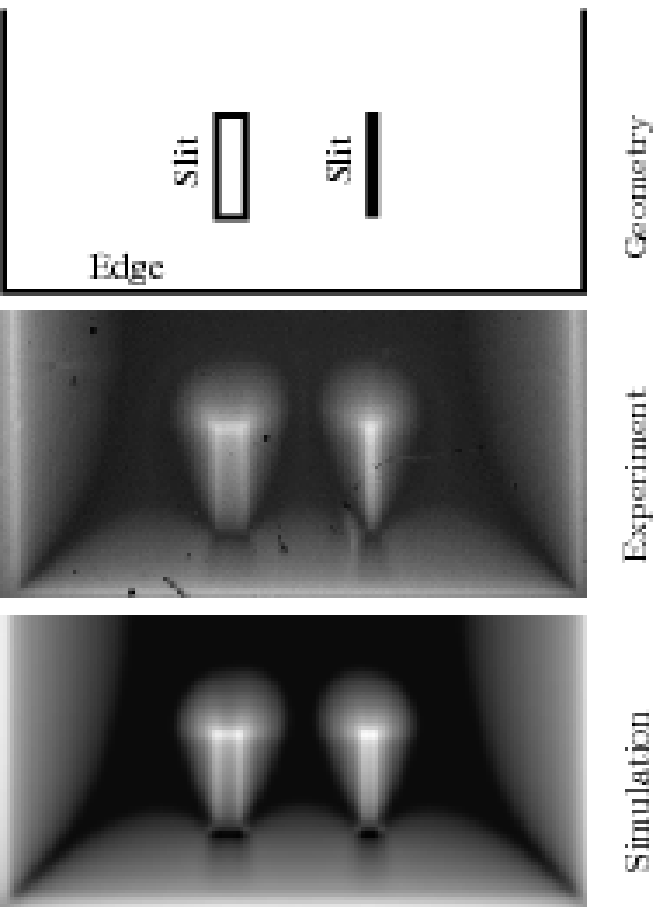, width=6cm}
  ~\vspace{0.5cm}~
  \epsfig{file=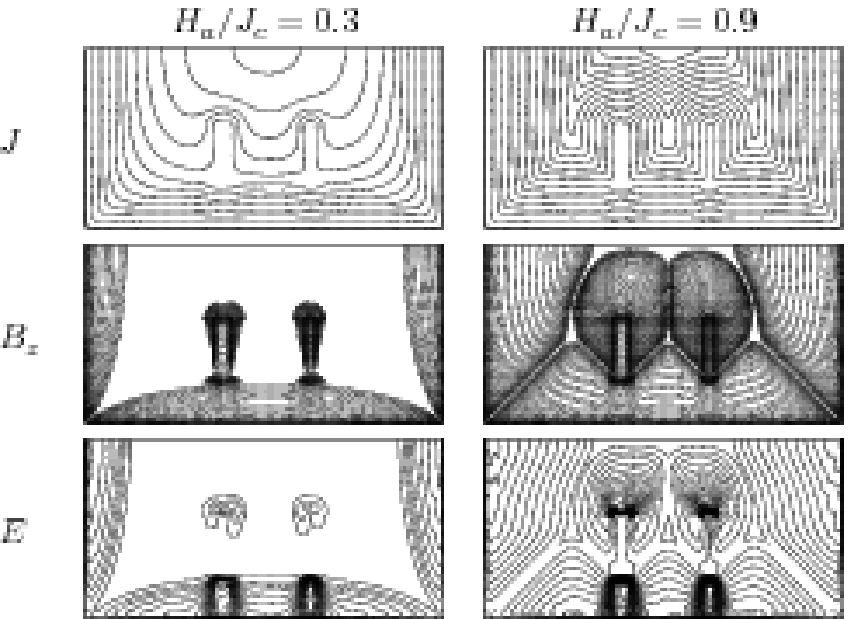, width=8.5cm}
  \caption{A square with two slits (only the lower half is shown).
    Top: Sample sketch, experimental magneto-optical image of  YBa$_2$Cu$_3$O$_x$ film,
    and simulated magnetic field distribution.
    Bottom: current stream lines, 
    $B_z$ and $E$ contour lines 
    at $H_a/J_c=0.3$ and~$0.9$, with $n=19$ and $\mu_0\dot H_a=\rho_0J_c/ad$.
    Note the strongly enhanced $J$ and $E$ between the slit and the edge 
    and the complicated set of $d$-lines at full penetration.
  }
   \label{fig:slits}
\end{figure}

Fig.~\ref{fig:flux-hole} shows the total flux
in a circular hole, $\Phi_h=\int_\text{hole}d^2r~B_z$, 
as a function of the applied
field $H_a$ for various distances between the hole and the edge.
In the beginning, $\Phi_h\approx 0$, until the main flux 
front is in contact with the hole. Then it starts to increase.
For high fields $\Phi_h$ grows almost linearly 
with $H_a$ at a universal growth rate determined by the hole area.
The linear rate is not just the case for small holes in strips, but has 
also been found for e.g. ring geometry.\cite{olsen07}
Note that for small fields $\Phi_h$ is close to,
but not exactly zero. 
The reason is the creation
of two additional flux fronts: one positive 
towards the flux-free region and one negative 
towards the edge, as also seen in Fig.~\ref{fig:strip-slices}. 
Only when integrating $B_z$ over a larger area that includes this
additional penetration one finds
that the total flux is exactly zero.\cite{brandt97} 
This integral is also a good consistency check of the 
boundary condition implementation, since a wrong 
value of $g^{(h)}$ tends to introduce a net, unphysical flux in the hole.

\section{Square with two slits}
\label{sec:slits}

This section presents results of simulations
of a square superconducting film with rectangular slits.
The sample geometry is chosen to reassemble one 
particular YBa$_2$Cu$_3$O$_{x}$ film
and comparison of the simulation with a magneto-optical image of the sample
is shown in Fig.~\ref{fig:slits}.
The experimental film thickness is $250$~nm, and side lengths are $2.5$~mm.
The two slits have been cut out with a laser.
Details of the film preparation can be found elsewhere.\cite{baziljevich96-2}

The experiment and simulations show a great similarity
both in large and in the details. The flux density 
is considerably enhanced everywhere along the slit edges, and reaches 
the maximal values at the upper 
corners. Our main result found for circular holes
holds true also for rectangular slits. Namely,
we again find a distinct dark region 
starting at the slit and widening towards 
the edge. It can be attributed to the over-critical 
current density in that region, which is clearly seen in 
the current stream-line plot.
A new result for slits is a slightly brightened regions near the upper corners
that appear due to concave current turns. A similar situation
arises in superconductors of some other shapes having concave 
corners, e.g. in crosses.\cite{schuster96}

There also exist a few minor discrepancies between 
flux distributions obtained in the
simulations and in the experiment of Fig.~\ref{fig:slits}.
The most notable is the details of the region of 
reduced $B_z$ at the side of slits close to the edge.
The values of $B_z$ appear to be less
in the simulation than in experiment. 
This might be caused by simplifications, like the
disregarded $B$-dependency of the material law or the 
simplification of using the sheet current in stead of 
the true current density.

\section{Summary}

We have proposed a new method for treating 
boundary conditions of non-conducting holes inside 
macroscopic, type-II superconducting films.
The key point is to reconstruct the at first unknown
$\dot B_z$ inside the holes, at each time step of the simulation.
The method is capable of handling any number of holes of arbitrary shape. 

The simulations of flux dynamics assuming a 
material law $E\sim j^n$ reproduce very well flux distributions 
observed by magneto-optical imaging in 
YBa$_2$Cu$_3$O$_x$ films, for circular holes 
as well as rectangular slits. In particular, 
they demonstrate a significant enhancement of current density 
in the region between a hole (slit) and the edge 
leading to a more intensive traffic of flux.
This region appears darker in magneto-optical images due to
a slight bending of current stream lines.

We thank C.~Romero-Salazar and Ch.~Jooss for fruitful discussions
and M.~Baziljevich for experimental data on Fig~\ref{fig:slits}.
This work was supported financially by The Norwegian Research
Council, Grant No. 158518/431 (NANOMAT) and by FUNMAT@UIO.

\appendix

\section{Numerical details}
\label{sec:numerics}

The simulations are run on a $N\times N$ square grid.
The creep exponent and the ramp rate are  
$n=19$ and $\mu_0\dot H_a=\rho_0J_c/ad$, a regime in which creep is low, 
but not negligible. Changing $n$ would only do quantitative changes 
to the results.
For small exponents the plateaus 
of current profiles, like Fig.~\ref{fig:j-slices}, would be 
less flat and there would also be more current 
compressed between the holes and the edge.

The main limiting factor of the simulations
is memory consumption since the kernel 
matrix $Q$, Eq.~\eqref{kernel-discrete},  has dimension $N^2\times N^2$.
The simulations are run with $N=100$ grid points, which
yields a kernel matrix of dimension $5000\times 5000$, when the
sample symmetry has been exploited.\cite{brandt95} 

The kernel $Q$ in Eq.~\eqref{kernel-discrete} depends explicitly on the sample shape.
Since the strip is infinite in the $y$-direction, 
$Q$ should be computed via an infinite sum over strip segments.
However a good approximation is achieved with only one 
segment on each side of the ``main'' strip. The strip segments further away
contain zero net current and the dipole like 
character means that they have a negligible effect.
A good accuracy of this approximation was checked by 
comparing 
the Meissner state width, $b$, obtained for very high $n$
with the analytical film Bean-model result,\cite{brandt93} 
$b=a/\cosh(\pi H_a/J_c)$.

The reconstruction of $\dot B_z$ inside the hole,
Eq.~\eqref{iteration-step},
need not use the full $Q$ from Eq.~\eqref{kernel-discrete}.
The best is to use a smaller kernel, $\tilde Q$,
also generated with  Eq.~\eqref{kernel-discrete},
but only including points inside the hole.
Fast convergence of  Eq.~\eqref{iteration-step} is achieve by ignoring 
currents at the hole perimeter,
which means that $\tilde Q$ should use $C(\mathbf r)=0$.

The most difficult numerical problem in our method
is the calculation of the electric field
at the boundary in Eq.~\eqref{constraint-hole}.
The electric field is given by the power 
law, Eq.~\eqref{material-law}, and is largely fluctuating between neighboring 
grid points. A stable way to handle this 
is to take the average of only the most significant values 
of $E$ and use $2\pi r$ and $\pi r^2$ for the 
hole circumference and area.


\begin{thebibliography}{33}
\bibitem[{not({\natexlab{a}})}]{note:mo-dark-bright}
\bibinfo{note}{What is dark and bright in MO-images depends on experimental
  setup. In this paper dark means low field and bright high field, which is the
  most common situation.}
\bibitem[{not({\natexlab{b}})}]{note:E-in-hole}
\bibinfo{note}{$E$ inside the hole cannot be found from the material law and is
  simply put to zero in the plots. The correct $E$ inside the hole must be
  found from Faraday's law.\cite{jooss06}}.
\end{thebibliography}

\end{document}